\documentclass[10pt,conference,letter]{IEEEtran}
\usepackage{graphicx} 
\usepackage{amsmath}
\usepackage{booktabs}
\usepackage{makecell}
\usepackage{color}
\usepackage[dvipsnames]{xcolor}
\usepackage{algorithm}
\usepackage{algpseudocode}
\usepackage{enumitem}
\usepackage{subcaption} 
\usepackage{amssymb}
\usepackage{pifont}
\usepackage{stfloats}
\usepackage{url}

\newcommand{\cmark}{\ding{51}}%
\newcommand{\xmark}{\ding{55}}%
\newcommand{\tool}{{ENFOR-SA}}

\title{\tool{}: End-to-end Cross-layer Transient Fault Injector for Efficient and Accurate DNN Reliability Assessment on Systolic Arrays
\vspace{-0.5em}
}

\author{
  \IEEEauthorblockN{
    Rafael B. Tonetto,
    Marcello Traiola,
    Fernando Fernandes dos Santos,
    Angeliki Kritikakou
  }

  \IEEEauthorblockA{
    Univ Rennes, CNRS, Inria, IRISA - UMR 6074, F-3500\\
    \{rafael.billig-tonetto, marcello.traiola, fernando.fernandes-dos-santos, angeliki.kritikakou\}@inria.fr
    \vspace{-6mm}
  }
}

\usepackage{fancyhdr}

\fancypagestyle{firstpage}{%
\fancyhf{}

 \lhead{This work has been accepted at IEEE VLSI Test Symposium (VTS) 2026}
 }

\begin{document}

\fontsize{9.99}{11.99}\selectfont

\bstctlcite{IEEEexample:BSTcontrol}
\maketitle

\begin{abstract}
Recent advances in deep learning have produced highly accurate but increasingly large and complex DNNs, making traditional fault-injection techniques impractical. Accurate fault analysis requires RTL-accurate hardware models. However, this significantly slows evaluation compared with software-only approaches, particularly when combined with expensive HDL instrumentation. In this work, we show that such high-overhead methods are unnecessary for systolic array (SA) architectures and propose \tool{}, an end-to-end framework for DNN transient fault analysis on SAs. Our two-step approach employs cross-layer simulation and uses RTL SA components only during fault injection, with the rest executed at the software level. Experiments on CNNs and Vision Transformers demonstrate that \tool{} achieves RTL-accurate fault injection with only 6\% average slowdown compared to software-based injection, while delivering at least two orders of magnitude speedup (average $569\times$) over full-SoC RTL simulation and a $2.03\times$ improvement over a state-of-the-art cross-layer RTL injection tool. \tool{} code is publicly available\footnote{\tool{} is available at \url{https://github.com/rafaabt/ENFOR-SA}}.

\end{abstract}

\thispagestyle{firstpage}

\section{Introduction}

Reliability assessment is crucial for evaluating DNN robustness to hardware faults, which can significantly degrade model accuracy. Fault injection is the primary method for identifying vulnerable components and designing fault-tolerance mechanisms, but modern DNNs have drastically expanded the fault space, making exhaustive evaluation impractical. For example, evaluating CIFAR-10 on ResNet-20 requires over 17 million simulations, translating to months of runtime~\cite{AnnachiaraDATE2023StatisticalFI}. Therefore, efficient fault injection techniques are essential to reduce evaluation time while preserving accuracy.

Producing realistic and statistically meaningful reliability results requires fault injection tools to account for fault masking effects across multiple abstraction levels~\cite{10.1109/ISCA52012.2021.00075}. High-level approaches (e.g., at the instruction or application level) offer rapid evaluation, but often neglect essential masking effects, and usually rely on simplified or incomplete fault models~\cite{10.1109/ISCA52012.2021.00075}. In contrast, low-level hardware simulations (e.g., RTL-based) provide high accuracy, but incur orders-of-magnitude longer execution times. These conflicting requirements highlight the need for scalable fault injection methodologies that balance accuracy and efficiency.
Cross-layer approaches attempt to bridge the gap between high-level speed and low-level accuracy by combining multiple simulation layers~\cite{DSN2021, thales, 10.1109/TC.2022.3184274,9251852}. Such methods have shown reasonable effectiveness for small and moderate-sized models (e.g., MLPs~\cite{10.1145/3195970.3196129,8203882,8368656}, LeNet-5~\cite{9116475,8203882,thales,11022806}, AlexNet~\cite{10.1145/3195970.3196129,8368656,e0d4219bd3c14e95b98c95b970c241c5,thales}, and small-dataset ResNets~\cite{9116475,e0d4219bd3c14e95b98c95b970c241c5}). Yet, accurately assessing the reliability of today’s highly complex DNNs, especially when deployed on increasingly common Systolic Array (SA) accelerators, requires much more aggressive solutions.
Even with existing cross-layer tools, evaluation times remain unacceptable for such modern workloads and architectures.

To address these challenges, we introduce \tool{}, a fast and accurate cross-layer fault-injection framework that integrates PyTorch-level DNN inference with efficient RTL fault injection. \tool{} is explicitly designed to support SA hardware accelerators, enabling scalable and realistic reliability assessment without requiring large infrastructures. As explained in detail in Section~\ref{sec:idea}, \tool{} performs high-speed inference at the Python level while enabling fine-grained, white-box RTL fault injection of SA components.
Extensive experimental results show that \tool{} significantly reduces evaluation time compared with full RTL fault-injection methodologies (average speedup of $569\times$, with a minimum of $198.66\times$), achieving an average slowdown of only 6\% compared to full-SW injection. In summary, our contributions are:

\begin{itemize}[leftmargin=*]

\item A fully integrated cross-layer framework that couples PyTorch-level inference with Verilator-based RTL fault injection for SA architectures. \tool{} maps intermediate tensors from DNN execution graph to RTL data paths, allowing for cycle-accurate RTL fault injection, while maintaining runtime close to software-level experiments.

\item A scalable methodology supporting diverse SA configurations and DNN workloads. \tool{} integrates SA implementations generated with Chipyard~\cite{chipyard}, enabling rapid exploration of microarchitectural parameters. \tool{} supports any pre-trained PyTorch model, allowing fault analysis across a range of networks and tasks.

\item A detailed evaluation against a state-of-the-art fault-injection framework. We conduct quantitative and qualitative comparisons, demonstrating that \tool{} overcomes key limitations of existing tools by achieving significantly faster fault-injection campaigns, while preserving RTL accuracy.

\item An open-source, end-to-end, and extensible framework enabling future research on DNN reliability. 




\end{itemize}

\section{Related Works}

\label{sec:related}

Fault injection frameworks have been proposed for reliability assessment dedicated to DNN workloads~\cite{feisu-survey,RuospoSurveyComputers,tallin_survey}. Table~\ref{tab:fi_papers} highlights representative fault-injection studies operating at different abstraction levels.


Simulation-based fault injections can be classified as Gate-Level (GL), Register-Transfer Level (RTL), microarchitecture, or software-level (SW). Speedup techniques have been proposed to accelerate GL simulation~\cite{10.1109/DAC63849.2025.11132928, 11071568}.
Other approaches adopt machine learning strategies that model GL behavior on a structured graph representation of the circuit's netlist~\cite{10390537, 9611337}. 
Fault analysis is performed by predicting a node's state from the graph representation. 
Although such methods maintain approximate accuracy for gate-level masking, none offer seamless end-to-end integration of DNNs with low-level accelerator representation. 

\begin{table}[b]
\vspace{-6pt}
\centering
\scriptsize
\setlength{\tabcolsep}{1.5pt}
\caption{Fault injection works supporting DNN workloads. 
}
\resizebox{\columnwidth}{!}{
\begin{tabular}{lccccc}
\toprule
\textbf{Paper} &
\textbf{\makecell{SA\\support}} &
\textbf{FI level} &
\textbf{\makecell{End-to-end \\framework}} & \textbf{\makecell{No instr.\\overhead}} \\
\midrule
\cite{10494993} (MRFI)                 & \xmark & SW & \cmark   & \cmark\\
\cite{10.1145/3295500.3356177} (BinFI) & \xmark & SW & \xmark   & \xmark\\
\cite{9151812} (PyTorchFI)             & \xmark & SW & \cmark   & \xmark\\
\hline
\cite{9978979} (LLTFI)            & \xmark & Architectural (ISA/IR) & \cmark &  \xmark\\
\hline
\cite{TAN2023251} (Saca-FI)            & \cmark & Microarchitecture & \cmark &  \cmark\\
\hline
\cite{10206919}                        & \cmark & RTL & \xmark &  \cmark \\
\cite{10.1145/3583781.3590226} (SiFI-AI) & \cmark & RTL & \cmark & \xmark \\
\cite{9251852} (FIdelity)                & \xmark & RTL (SW model) & \cmark & \cmark\\
\cite{10508925} (SAFFIRA)                & \cmark & RTL (URE equations) & \cmark & \cmark\\
\cite{10154505}                & \cmark & RTL & \xmark & \xmark\\
\hline
\cite{10.1109/DAC63849.2025.11132928} (EPICS) & \xmark & GL & \xmark &\cmark \\
\cite{11071568} & \xmark & GL & \xmark &\xmark \\
\cite{10390537} & \xmark & GL (fault model) & \xmark &   \cmark\\
\cite{9611337} & \cmark  & GL (fault model) & \xmark &  \cmark \\
\cite{9791408} (HDFIT)   & \cmark & GL (RTL also possible) & \xmark & \xmark \\ \hline
\textbf{\tool{}} (ours) & \cmark & RTL & \cmark  & \cmark \\
\bottomrule
\end{tabular}
}
\label{tab:fi_papers}
\end{table}

At RTL, the approaches of~\cite{10206919,10.1145/3583781.3590226} propose injection methodologies for the Gemmini systolic array \cite{10.1109/DAC18074.2021.9586216}. In \cite{10206919}, although accurate, the simulation is performed entirely at RTL and no simulation time improvement methodology is provided. Thus, fault analysis tends to be highly constrained in terms of the input space, the number of evaluated DNN layers, and the hardware configurations. Alternatively, \cite{10.1145/3583781.3590226} proposes a hybrid approach in which only the code portion subject to faults executes in detailed RTL models. However, it requires heavy RTL code annotations - in the Hardware Description Language (HDL) - for each individual fault target (fault position, cycle, or hardware-specific conditions must be annotated in HDL), and no further speedup methodology is provided. 
FIdelity~\cite{9251852} proposes an error model framework for fast and accurate resilience analysis of DNN accelerators by incorporating hardware semantics to approximate RTL effects. 
While it provides an end-to-end framework achieving near-RTL accuracy, it is not a full RTL replacement and does not reproduce the exact faulty values produced by some faults (e.g., local-control FFs), replacing them with random values. Finally, it has been validated on NVDLA-like accelerators rather than on systolic arrays.
%
%
SAFFIRA~\cite{10508925} proposes an end-to-end HW-aware fault injection method using a SA model derived from Uniform Recurrent Equations (URE), capturing spatial and temporal behavior. Although it is much faster than RTL, it sacrifices accuracy by missing RTL-only effects (e.g., control and interconnects). Another study~\cite{10154505} presents an RTL–SW cross-layer framework to balance speed and accuracy. Still, it requires heavy RTL instrumentation, relies on the less-used N2D2 framework, lacks full end-to-end support, and remains time-consuming because it simulates full convolution channels in RTL.

Micro-architectural fault models for SAs have also been proposed \cite{TAN2023251}. While such approaches suffice for fault analysis on storage-based components (e.g., scratchpad memories, input buffers, etc.), they lack hardware representation of key components of SAs that also influence fault characterization, such as clock events modeling, pipeline registers, and control-related signals. Sec. \ref{sec:results} shows how lower-level fault analysis is more suitable to guide selective protection mechanisms.
 
SW-based solutions~\cite{10494993, 10.1145/3295500.3356177,9151812} or ISA-level approaches \cite{9978979} have been proposed. Although these approaches are faster than HW simulation, they come at the cost of low accuracy, as such fault models flip bits directly in the SW variables (e.g., the layer outputs or weights) or intermediate representations (e.g., ISA registers), and do not take into account how tensors and operations are mapped to HW during computation. This results in faulty output layers, such as single-bit errors, that do not accurately reflect realistic scenarios. Indeed, when a HW model is used, a single bit flip in a register can lead to multiple bit errors in the output.

Among all works shown in Tab. \ref{tab:fi_papers}, \textit{HDFIT}~\cite{9791408} is the only open-source tool fully comparable to our approach, sharing the same fault abstraction level and the ability to map fault instrumentation to an SA architecture. 
The HDFIT tool provides a fine-grained methodology for instrumenting hardware descriptions of arbitrary designs with fault-injection capabilities in Verilator \cite{verilator}. 
HDFIT was initially proposed for DNN reliability assessment and was also used for fault analysis of HPC workloads~\cite{10018868}, where customized matrix multiplication kernels are offloaded to a gate-level SA simulation. HDFIT can be used with arbitrary hardware descriptions, supporting gate-level netlists or RTL designs. One key shortcoming, however, is that \textit{HDFIT relies on instrumenting every combinational and sequential assignment in the HDL}. We claim that this is unnecessary for regular RTL designs. 
Finally, HDFIT does not provide an end-to-end integration with widely used frameworks (e.g., PyTorch). However, thanks to its low-level, API-based approach (oneDNN), we integrated HDFIT-instrumented verilated designs at the RTL level (i.e., not at the gate level) into our framework.
As described in the next section, \tool{} achieves faster RTL injection time by avoiding such high-overhead HDL instrumentation.

\section{Proposed Reliability Assessment Framework}
\label{sec:idea}


\subsection{Verilator Fault Injection Approach}

\begin{figure}[t!]
\centering 
\includegraphics[width=0.87\columnwidth]{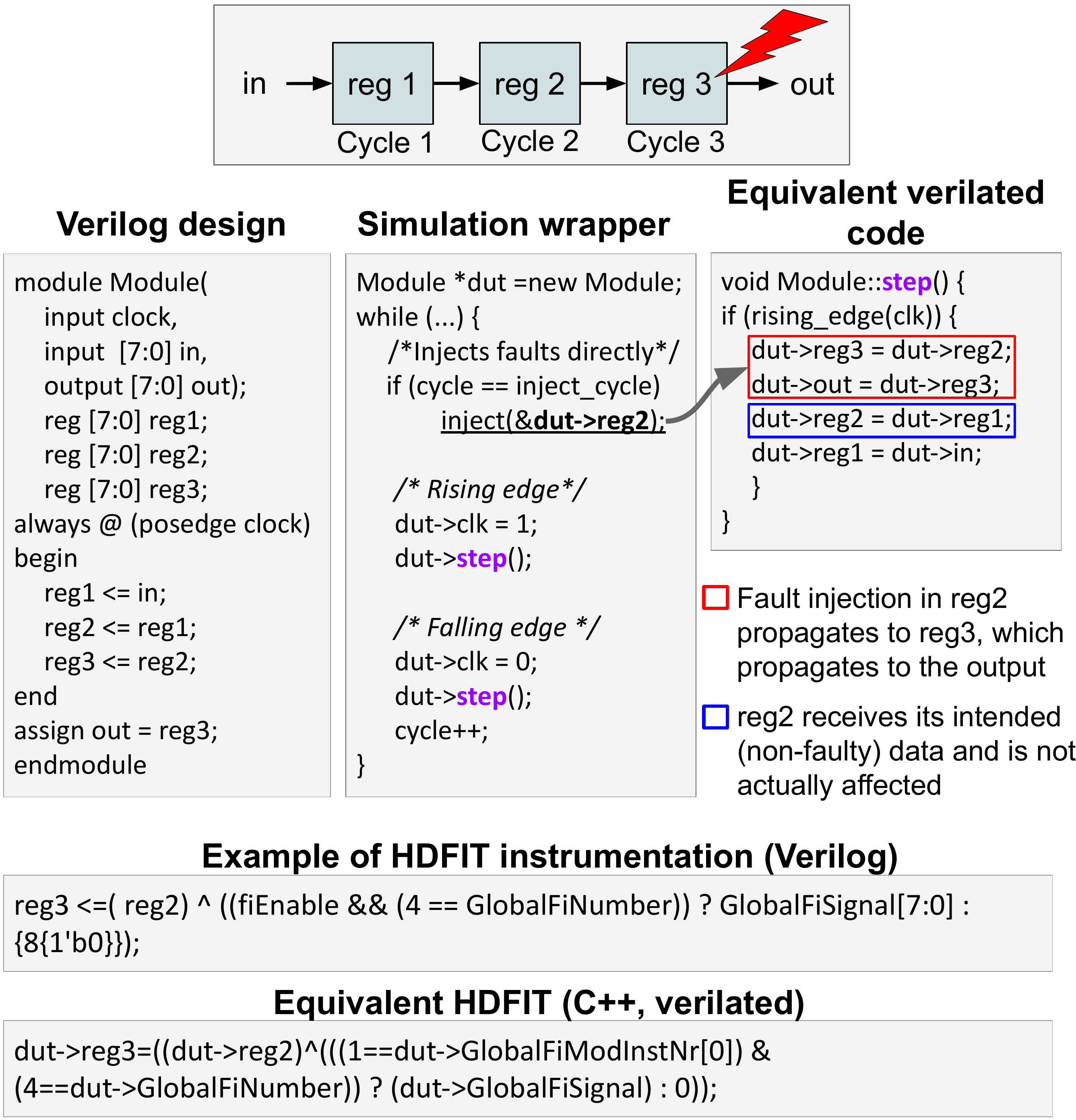}
\caption{
Example of Verilog and equivalent verilated codes for a chain of register assignments: \textbf{\tool{}} injects in reg3 by targeting the reg2 variable; 
\textbf{HDFIT} requires instrumenting each assignment, resulting in additional operations per cycle, even when a single signal is injected (e.g., an 8$\times$8 mesh has 632 assignments, all instrumented). 
}
\vspace{-1.5em}
\label{fig:chain}
\end{figure}

We propose an alternative solution for Verilator-based fault injection. Unlike previous works, such as \cite{10.1145/3583781.3590226,9791408}, our solution does not require HDL instrumentation, thereby avoiding the associated runtime overhead. Fig. \ref{fig:chain} illustrates the general idea with an example. It shows the Verilog code for a simple register assignment chain and the equivalent C++ (verilated) code used in simulation. 
Verilator generates a C++ RTL model that is functionally equivalent to the original Verilog description. To test the circuit, the user sets the input signals and advances the simulation by toggling the clock and invoking the model’s \textit{step()} function to execute a complete evaluation step. To preserve the semantics of Verilog’s register updates, Verilator inverts the order of register assignments, as the figure shows, ensuring that each register’s update cycle remains consistent with the hardware behavior. This transformation enables a simple yet effective fault-injection strategy: instead of modifying the HDL (e.g., similar to how HDFIT operates), one can inject into the register’s data \textit{source} by acting on the simulation wrapper code 
achieving the same effect with a cleaner, more controllable, and free from HDL instrumentation overhead. 
The same idea of Fig.\ref{fig:chain} applies to SA architectures, where PEs propagate signals to neighbours.
\begin{figure}[b!]
\vspace{-10pt}
\centering 
\includegraphics[width=0.87\columnwidth,trim={9 6 8 2}, clip]{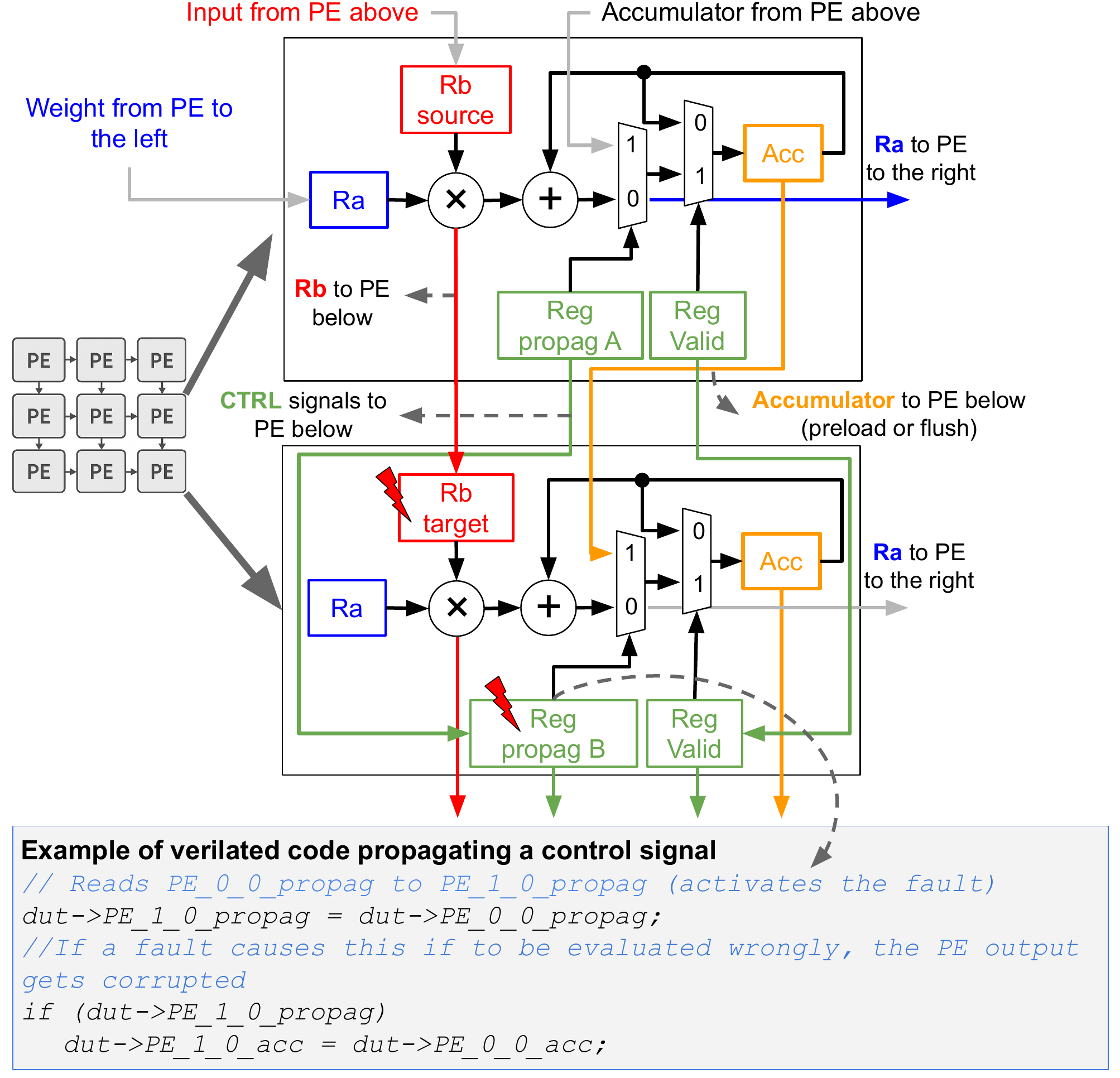}
\caption{The Gemmini PE's architecture (Simplified. Output stationary). \tool{} can inject in all signals inside the PEs.}
\label{fig:pe}
\end{figure}
Hence, we integrate our injection approach with the Gemmini SA \cite{10.1109/DAC18074.2021.9586216}. In Gemmini, we can target any signal inside the Processing Elements (PEs), as shown in Fig. \ref{fig:pe}. In particular, we can inject all pipelined registers used to propagate the inputs and weights, accumulators,  and local control signals (\textit{Propag} and \textit{Valid}). We provide a brief analysis of the effects of errors in these bits in Sec. \ref{sec:results}. Note that, for highly regular architectures, such as an RTL SA, injection points can be instrumented by tracking the source-register pointers associated with each PE signal. Fault injection is then performed by randomly flipping bits in these pointers (i.e., the \textit{inject()} function), effectively corrupting the data sources feeding the PE without modifying the register structures themselves or requiring high-overhead HDL instrumentation. Following the example shown in Fig. \ref{fig:pe}, fault injections in the register \textit{Rb target} (lower PE) can be simulated by targeting \textit{Rb source} (upper PE). Note that \textit{Rb source} is not actually affected due to the way Verilator layouts the code by inverting the assignment order: \textit{Rb target} is assigned first, copying the contents of \textit{Rb source} (fault injected). \textit{Rb source} is assigned afterwards, receiving its intended data from its preceding PE. Due to the nature of the SA, we can apply the same approach to all signals within the PEs, including the control signals, which are also propagated through the array. For example, to inject in \textit{Reg propag B}, one can target \textit{Reg propag A}, in the upper PE. The lower part of Fig. \ref{fig:pe} shows a code snippet of a real verilated case for the propagation of a control signal, e.g., the \textit{PE\_1\_0} propagate register receives data from the PE above it (\textit{PE\_0\_0}). 



\subsection{Mesh Isolation for Faster Cross-layer Fault Injection}

\begin{figure}[b!]
\vspace{-10pt}
\centering 
\includegraphics[width=0.85\columnwidth,trim={2 3 2 2}, clip]{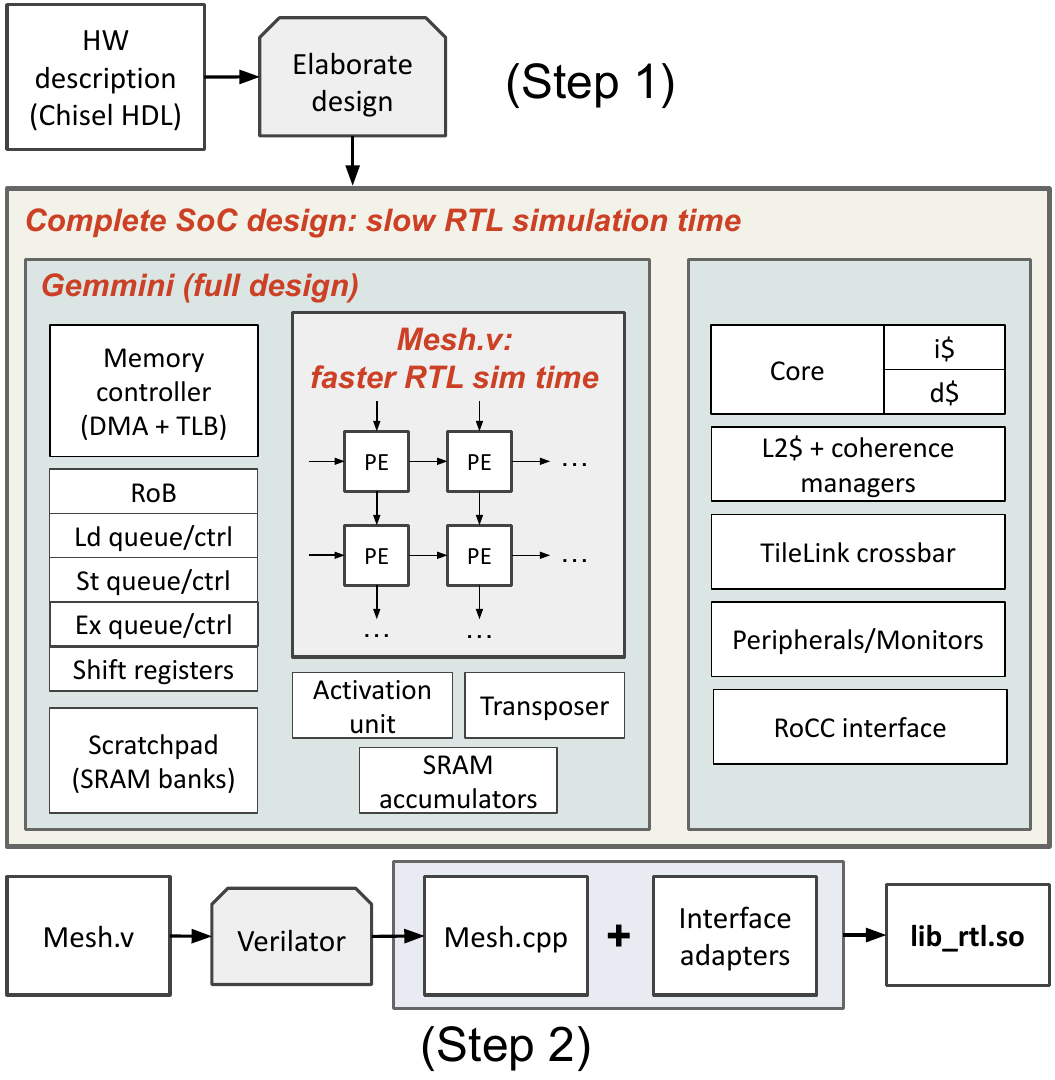}
\caption{
In step 1, we generate the original SoC system; in step 2, we extract the Mesh module and couple it to interface adapters that emulate major hardware blocks required for systolic simulation (e.g., shift registers, transposers). A library is generated and used at runtime.}
\label{fig:dt}
\vspace{-4mm}
\end{figure}


Fig. \ref{fig:dt} shows an abstract view of a conventional heterogeneous SoC design from Chipyard~\cite{chipyard}, consisting of a core, cache memories and a Gemmini SA. Typically, design exploration is conducted by specifying the relevant hardware parameters, elaborating the design (Step 1), and then mapping it to FPGAs or using Verilator~\cite{verilator} to simulate the design in a host machine. Our framework works on the basis of Verilator. 
Though verilator has been shown to generate well optimized code for RTL simulation - even surpassing commercial tools in terms of simulation performance~\cite{9218632,10.1145/3545008.3545091}, the high complexity of a complete SoC model imposes prohibitive overheads for large-scale fault injection campaigns.


\subsubsection{Compilation time}
As shown in Fig. \ref{fig:dt}, the reference accelerator is (loosely) coupled to a core, together with the full memory hierarchy, crossbars, peripherals, and the interfaces to communicate with the SA. On the accelerator side, a full SA architecture is available, with scratchpad memories, a DMA engine, control-related circuitry, an activation unit, and the \textit{Mesh} unit, which comprises arrays of Multiply-Accumulate (MAC) units for matrix multiplication. Gemmini provides both output-stationary (OS) and weight-stationary (WS) execution modes.
Our goal is to perform fault analysis on the SA by capturing only Mesh-related computation (i.e., the \textit{Mesh.v} block) and its masking properties. This unit comprises pipelined MAC units and all relevant RTL signals for fault injection. Essential signals are the PE input and output (accumulator) registers, and valid/propagate control signals. Our approach consists of a \textbf{compilation} phase, where we generate the SoC (Step 1) and isolate the Mesh unit (Step 2) to avoid RTL simulation of other HW blocks. This phase is done only once per HW configuration, and is transparent w.r.t. DNN models.

\begin{figure*}[t!]
\centering 
\includegraphics[width=0.78\textwidth]{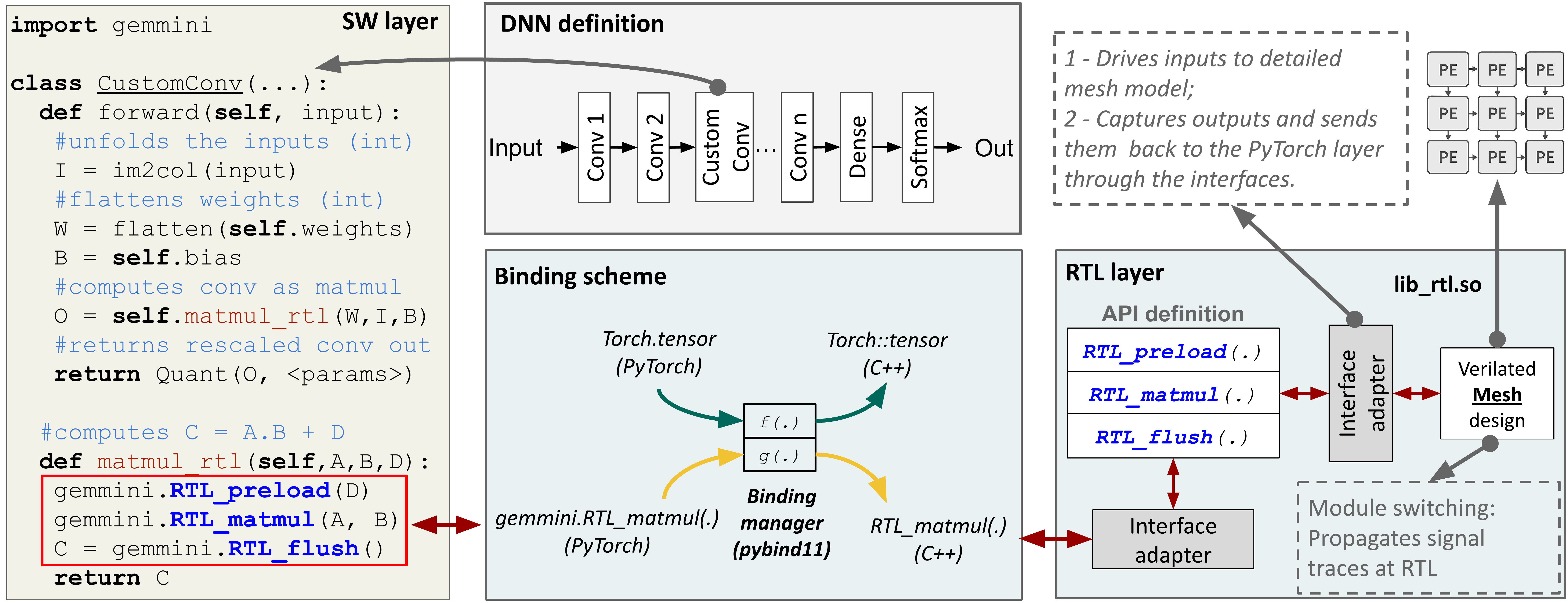}
\caption{Runtime cross-layer call stack. A lightweight model instrumentation offloads the (per-tile) matmul kernel (e.g., convolutions or an attention blocks) down to an RTL SA module to perform matrix multiplication. The SA result is flushed from the internal PE accumulators and sent back to the PyTorch layer through fast interfaces.}
\label{fig:rt}
\vspace{-4mm}
\end{figure*}


\subsubsection{Runtime}
Fig. \ref{fig:rt} shows the \textbf{runtime} phase of our tool. We use interface adapters to drive PyTorch tensors to the Mesh unit. By bypassing hardware components irrelevant to fault analysis, our platform delivers significant speedups over conventional full-SoC simulation (details in Sec. \ref{sec:results}).
The \tool{} tool enables fault injection into pretrained PyTorch CNN models~\cite{torchvision} and Vision Transformer (ViT) models~\cite{10378350}. For CNNs, fault injection is performed by hooking the forward pass of target convolutional layers: to map this type of layer to Gemmini, convolutions are expressed as matrix multiplications by using the im2col procedure. For the ViT models, we target matmul-related tasks inside the attention blocks and the classification layer. 
As Fig.~\ref{fig:rt} shows, at runtime, the forward pass of a target layer is redirected to a custom implementation that binds to Gemmini mesh. For a single transient fault, only a single activation tile 
and one from the weight tensors is extracted and offloaded to the accelerator. Cross-layer simulation is achieved by using a low-cost binding approach (pybind11) to bind the PyTorch-level tensor tiles to C++, which is then converted to native data types for direct offloading to the Mesh for matrix multiplication in Verilator. To perform the matrix multiplication, the complete Mesh unit is simulated, and a single transient fault is injected into a given target MAC unit during computation. Lastly, the matrix multiplication result is flushed from the mesh (in OS mode) and mapped back to the PyTorch layer's output.


\section{Experimental results}

\subsection{Experimental setup}


\begin{table}[b!]
\centering
\scriptsize
\caption{Evaluated quantized models.
}

\label{tab:models}
\begin{tabular}{lcc}
\toprule
\textbf{
\makecell{Quantized\\model
}
} & \textbf{\makecell{Accuracy\\(Top-1)}} & \textbf{Parameters} \\ 
\midrule
MobileNetV2			& 71.60\%	& 3.50M\\
DeiT-T			    & 72.24\%	& 5.00M\\
GoogLeNet			& 69.8\%	& 6.60M\\
ShuffleNetX20	    & 75.3\%	& 7.40M	\\
ResNet18			& 69.4\%	& 11.7M\\
DeiT-S 				& 80.1\%	& 22.0M\\
ResNet50			& 80.2\%	& 25.6M\\
InceptionV3		    & 77.1\%	& 27.2M\\
ResNeXt64			& 82.8\%	& 83.5M	\\
ResNeXt32			& 82.5\%	& 88.8M\\
\bottomrule
\end{tabular}
\vspace{-3mm}
\end{table}

We experiment with the pre-trained quantized models shown in Tab.~\ref{tab:models}~\cite{torchvision,ivit}. These models are optimized for reduced-precision integer arithmetic (int8) and can be seamlessly integrated with state-of-the-art accelerators such as Gemmini. Our experimental setup uses a 96-core AMD EPYC 7443 server processor @2.8GHz for the CNNs, as the quantized CNN models do not support CUDA. With the ViT models, we were able to utilize an NVIDIA A40 GPU. We use a subset of 20 input batches, each with 32 inputs, randomly selected from the ImageNet validation set. Our experiments aim to 1) compare the achievable performance improvements w.r.t the HDFIT tool, and 2) show that \tool{} achieves RTL accuracy with performance comparable to SW-based injections. Thus, we experiment with three fault injection scenarios: 1) our \tool{} methodology, 2) \textit{HDFIT} instrumentation approach; and 3) software-only fault injection, used as a baseline for injection time comparison. 
For RTL performance evaluation, we consider varying array configurations. For injection experiments, we used an 8$\times$8 OS config (referred to as DIM8). For both SW and RTL, we inject 500 \textbf{transient faults} per layer per input, totaling nearly $42M$ faults and ensuring statistical significance according to~\cite{AnnachiaraDATE2023StatisticalFI}. From the fault injection campaigns, we derive the Architectural Vulnerability Factor (AVF) \cite{1253181} of the array, based on the percentage of \textit{critical} inferences across all injection trials. We consider a fault as \textit{critical} if it causes the model's Top-1 classification label to diverge from the golden (fault-free) classification.

\subsection{Evaluation Results}
\label{sec:results}


We compare the mean cycle time of \tool{} to the \textit{HDFIT} case (both Mesh-only). Tab. \ref{tab:cycle_time} shows the mean cycle time, averaged after 1 million simulation cycles (i.e., 1M raw calls to \textit{$dut\rightarrow step()$}), for both approaches.
We used the same SA configurations for both HDFIT and \tool{} at the RTL level, but instrumented HDFIT HDL for fault injection, as described in the original HDFIT approach. HDFIT's heavy instrumentation incurs considerable runtime overhead compared to non-instrumented HDL. Thus, \tool{} achieves an average performance improvement of $2.66\times$.

\begin{table}[b!]
\vspace{-15pt}
\centering
\caption{Mean cycle time for the explored array sizes.}
\label{tab:cycle_time}
\begin{tabular}{lrrr}
\toprule
\textbf{Array Size} & \textbf{\makecell{\tool{}\\(mesh only)}} & \textbf{\makecell{HDFIT\\(mesh only)}} & \textbf{Improvement} \\ 
\midrule
DIM4	& 0.084$\mu$s	& 0.245$\mu$s	& 2.91$\times$ \\
DIM8	& 0.403$\mu$s	& 1.189$\mu$s	& 2.95$\times$ \\
DIM16	& 1.859$\mu$s	& 5.788$\mu$s	& 3.11$\times$ \\
DIM32	& 16.73$\mu$s	& 39.65$\mu$s	& 2.36$\times$ \\
DIM64	& 100.1$\mu$s	& 199.9$\mu$s	& 1.99$\times$ \\
\bottomrule
\end{tabular}

\end{table}

\noindent{\bf Matrix multiplication performance:}
Tab. \ref{tab:matmul_time} shows the mean time taken to perform matrix multiplications ($C=A \cdot B + D$) with varying array sizes (averaged after 1k matmuls) for both \tool{} and \textit{HDFIT} cases. The time taken comprises the cumulative steps of all phases of systolic computation, accounting for: time for preloading a bias matrix (D) to the PE accumulators, propagating the input matrices through the array (A and B) and performing MAC computations, and flushing the PE output into a result matrix (C). On average, we outperform HDFIT by a factor of $2.19\times$.

\noindent{\bf{Case study - ResNet50 performance:}}
We assess the performance improvements of \tool{} w.r.t the full-SoC design by measuring the time taken for a full forward pass on the ResNet50's first convolution layer, shown in Tab.~\ref{tab:resnet50_time}. We simulate the layer in three cases: (i) the \textit{full-SoC}, the original Chipyard Gemmini accelerator, consisting of the complete chip design, encompassing a RISC-V Rocket core + the full Gemmini SA, (ii) the \textit{Mesh-only HDFIT} design, and (iii) \tool{}. All approaches run completely on Verilator.
\noindent As expected, simulating only the Mesh design has significant performance improvements w.r.t. the full-SoC simulation. Note that the performance improvement decreases as the array size increases. As all configurations differ only in the number of PEs (sharing the same configurations for all other parameters), the proportion of Mesh signals increases with larger DIM sizes, thereby reducing the effectiveness of Mesh-only simulation relative to the whole chip. Moreover, \tool{} improves upon the full-SoC by at least \textit{two orders of magnitude}. HDFIT imposes significant overhead when compared to \tool{}. Across all configurations, \tool{} outperforms HDFIT by an average of $2.03\times$ in terms of simulation performance.

\begin{table}[b!]
 \vspace{-15pt}
\centering
\caption{Mean matmul time for the explored array configs. 
}
\label{tab:matmul_time}
\begin{tabular}{lrrr}
\toprule
\textbf{Array Size} & \textbf{\makecell{\tool{} \\ (mesh only)}} & \textbf{\makecell{HDFIT\\(mesh only)}} & \textbf{Improvement} \\ 
\midrule

DIM4	& 0.006ms	& 0.014ms	& 2.08$\times$ \\
DIM8	& 0.040ms	& 0.087ms	& 2.15$\times$ \\
DIM16	& 0.248ms	& 0.669ms	& 2.69$\times$ \\
DIM32	& 4.607ms	& 9.328ms	& 2.02$\times$ \\
DIM64	& 48.97ms	& 98.02ms	& 2.00$\times$ \\

\bottomrule
\end{tabular}
\end{table}

\begin{figure}[b!] 
\vspace{-8pt}
\centering
\subfloat[Targeting the control signals.]{\label{fig:res:ctrl}
\includegraphics[width=0.49\columnwidth,trim={30 10 5 42}, clip]{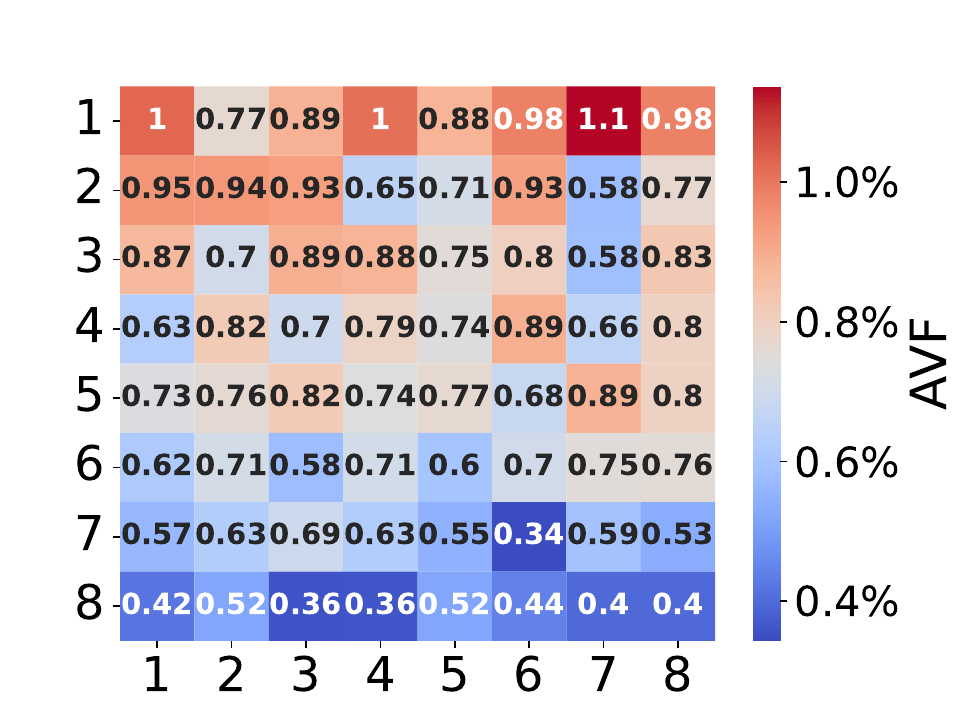}}
\subfloat[Targeting the registers holding weights (left to right propagation)]{\label{fig:res:data}
\includegraphics[width=0.49\columnwidth,trim={30 10 5 42}, clip]{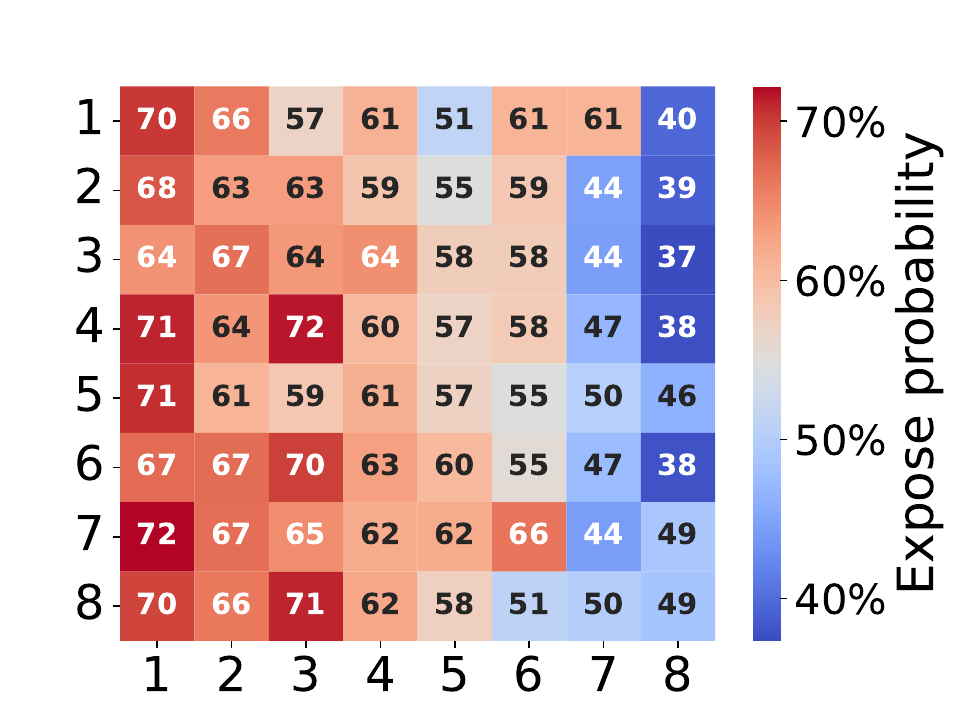}}
\caption{ReNet50: a) AVF for control signals. b) Fault exposure probability for data registers. The config is an 8$\times$8 SA (OS).}
\label{fig:res_data_ctrl}
\end{figure}

\noindent{\bf{Case study - ResNet50 injection:}}
To demonstrate the capabilities of our approach, we perform cross-layer fault injections by offloading a single matmul to RTL (to be injected). We consider faults in control-related signals of the Gemmini SA, which are typically available only in RTL representation. In Gemmini, each PE has two local control signals: 1) a \textit{propag} signal is used to propagate the data in the PE accumulators to the next PE below (used for preloading bias matrices and flushing the results out of the SA), and 2) a \textit{valid} signal is asserted for every new incoming valid data in each PE. Fig. \ref{fig:res:ctrl} shows the AVF for each PE when control signals are affected. When \textit{valid=1}, the PE will perform a MAC operation using the inputs for that cycle. If a fault sets \textit{valid=0}, the PE will not update its accumulator for one cycle, corrupting the PE output. During output-stationary computation, the \textit{propag} bit must remain fixed at 0 (it is set to 1 only during preloading and output flushes). Consequently, a fault that erroneously asserts this bit to 1 during computation forces the PE accumulator to receive propagated accumulator data from the PE above for one cycle. Since this bit is forwarded to downstream PEs, the resulting corruption propagates through the entire column (affecting the target PE and all PEs below it), thus making the upper rows more critical. This insight is particularly relevant for evaluating selective protection mechanisms at the PE level, where a low-level architectural representation is necessary.

\begin{table}[t!]
\centering
\caption{Mean time for a full forward pass of the ResNet50's 1\textsuperscript{st} convolution layer. All cases have RTL accuracy.}
\label{tab:resnet50_time}
\setlength{\tabcolsep}{2pt}
\resizebox{\columnwidth}{!}{
\begin{tabular}{lc|cc|cc}
\toprule
\textbf{\makecell{Array\\Size}} & \textbf{\makecell{\tool{}\\(mesh only)}} & \textbf{\makecell{Full\\SoC}} & \textbf{\makecell{\tool{}\\vs Full SoC}}&\textbf{\makecell{HDFIT\\(mesh only)}} & \textbf{\makecell{\tool{}\\vs HDFIT}}  \\ 
\midrule
DIM4 & 19.8s & 06h21min & 1155.6$\times$ & 31.4s & 1.58$\times$ \\
DIM8 & 5.86s & 01h16min & 785.23$\times$ & 10.6s & 1.82$\times$ \\
DIM16 & 3.75s & 28min55s & 462.55$\times$ & 9.51s & 2.53$\times$ \\
DIM32 & 7.35s & 29min24s & 240.33$\times$ & 17.2s & 2.34$\times$ \\
DIM64 & 17.4s & 57min46s & 198.66$\times$ & 32.7s & 1.87$\times$\\
\bottomrule
\end{tabular}
}
\vspace{-1.5em}
\end{table}




We complement our analysis by injecting faults into the registers holding the DNN weights, which in our configuration flow horizontally across the SA from left to right (though in OS mode, weights may also flow vertically from top to bottom). Faults in these registers may be masked when multiplied by zero-valued activations. Figure \ref{fig:res:data} reports the probability that a fault in these registers becomes \textit{exposed} to the SW layer, i.e., the probability of not being masked, thus corrupting a layer's output. Notably, the first columns are more critical: faults occurring in “earlier” PEs have more opportunities to propagate to subsequent units (i.e., the fault tends to be more ``reused'' along the row), thereby reducing the likelihood of fault masking within the array. Note that the software layer may mask exposed faults due to inherent DNN noise robustness, so high exposure probabilities do not necessarily translate into high model AVF.



Table~\ref{tab:inj_time_vs_sw} shows the estimated AVF (computed at RTL) for DNNs. For each DNN, we also estimate the Program Vulnerability Factor (PVF) obtained from SW-only injections. The PVF only encapsulates SW-level masking properties \cite{10.1145/1366224.1366225} and does not account for HW-level masking. This yields pessimistic vulnerability estimates. Overall, our mean PVF estimation is $5.3\times$ higher than the mean AVF. Although this observation is not new \cite{10.1145/1815961.1816023}, we leverage it here to illustrate the gap between HW-aware and HW-agnostic fault injection, further justifying the importance of accurate analysis.

\begin{table}[t!]
\centering
\caption{Injection time and AVF/PVF vulnerability factors. 
}
\label{tab:inj_time_vs_sw}
\setlength{\tabcolsep}{5.4pt}
\begin{tabular}{lccccc}
\toprule
\textbf{
Model
} & \textbf{\makecell{SW\\(inputs)}} & \textbf{\makecell{\tool{}\\(RTL)}} & \textbf{\makecell{Slowdown}} & \textbf{PVF}$^*$& \textbf{\makecell{AVF}$^*$
}\\ 
\midrule
MobileNetV2    	& 2.44h	& 2.80h	& 14.7\% & 2.67\% & 1.00\%\\
DeiT-T	        & 4.76h	& 4.93h	& 3.57\% & 9.95\% & 0.91\%\\
GoogLeNet	    & 2.73h	& 3.10h	& 13.5\% & 0.97\% & 0.61\%\\
SuffleNetX20	& 12.1h	& 12.6h	& 4.13\% & 2.11\% & 0.60\%\\
ResNet18	    & 1.66h	& 1.80h	& 8.43\% & 1.76\% & 0.64\%\\
DeiT-S	        & 7.93h	& 8.21h	& 3.53\% & 9.02\% & 0.85\%\\
ResNet50	    & 6.64h	& 7.00h	& 5.42\% & 0.97\% & 0.34\%\\
InceptionV3	    & 7.90h	& 8.00h	& 1.26\% & 0.76\% & 0.29\%\\
ResNeXt64	    & 24.0h	& 24.6h	& 2.71\% & 1.02\% & 0.19\%\\
ResNeXt32	    & 25.4h	& 26.0h	& 2.67\% & 0.60\% & 0.11\%\\
\midrule
Mean	        & 9.55h & 9.91h & 6.00\% & 2.98\% & 0.55\%\\
\bottomrule


\multicolumn{5}{l}{\footnotesize $^*$percentage of critical inferences}\\
\end{tabular}
\vspace{-15pt}
\end{table}

\noindent{\bf{Injection time evaluation:}}
Tab. \ref{tab:inj_time_vs_sw} shows the total injection time for all models. We compare the SW-only injection time, where injections are performed in SW variables (with no RTL modules involved) against our cross-layer procedure (faults at RTL). Although faster, SW injection does not capture any HW-related masking properties, thus providing less accurate results, as previously shown.  

Other HW-agnostic tools, such as LLTFI (ISA-level error modeling) \cite{9978979}, suffer from instrumentation overheads of up to 117\% (mean 68\%). We take a different direction by offloading only a single tile for RTL injection, resulting in a relatively modest maximum overhead of 14.7\% (mean 6\%). This enables us to deliver RTL-accurate results with lower overhead than previous (less accurate) ISA-level approaches. Overall, our method bridges the performance-accuracy gap in error analysis, achieving RTL fidelity while maintaining injection times comparable to SW-based techniques.

\noindent{\bf{Accuracy validation against HDFIT~\cite{9791408}:}}
We validate \tool{} accuracy by comparing it with the fully instrumented HDFIT method. For both \tool{} and HDFIT, we perform random matmuls while injecting faults into the PE signals. The same input matrices, fault locations, and injection cycles are used for both methods. We then compare the resulting faulty output matrices and confirm that \tool{} and HDFIT produce identical results. This demonstrates that our approach reproduces HDFIT’s injection behavior with no loss in accuracy while enabling significantly faster analysis.

\section{Conclusions and Future Work}
We propose an accurate and efficient end-to-end transient fault injection methodology that tightly integrates DNN workloads with RTL-level SA hardware models. To reduce RTL simulation time, we introduce a non-intrusive injection method and 
target only the Mesh component of the SA. Our approach uses cross-layer simulation by mapping to the RTL level only the PyTorch tensors to inject, while executing the remaining simulation in software. We achieve a speedup of at least two orders of magnitude compared to full-SoC RTL simulation, with an average speedup of $569\times$ and a runtime similar to the full-SW approach (only a 6\% average slowdown). Additionally, we outperform a state-of-the-art instrumentation tool by an average of $2.03\times$ in injection performance, all while maintaining RTL accuracy.

\section*{Acknowledgment}
This work was supported by the French National Research Agency through the RE-TRUSTING project ANR-21-CE24-0015 and the FASY project ANR-21-CE25-0008-01.


\bibliographystyle{IEEEtran}
\bibliography{biblio.bib}

\end{document}